\documentclass[twocolumn,pre]{revtex4}
\usepackage{graphicx}
\usepackage{amsmath}
\usepackage{amssymb}
\usepackage{epsfig}

\newcommand{\beq}{\begin{equation}}
\newcommand{\eeq}{\end{equation}}
\newcommand{\beqa}{\begin{eqnarray}}
\newcommand{\eeqa}{\end{eqnarray}}

\newcommand{\mat}[1]{{\cal {#1}}}

\def\d{d}
\def\a{a}
\def\b{b}
\def\c{c}

\def\D{\Delta}

\def\e{\epsilon}
\newcommand{\Tr}{{\rm Tr}}

\newcommand{\z}{\hat z}
\newcommand{\y}{\hat y}
\newcommand{\vv}{\hat v}
\newcommand{\lam}{\hat \lambda}
\newcommand{\vi}{\hat v_i}
\newcommand{\Di}{\hat \Delta_i}

\newcommand{\sumi}{\sum_i^N}

\newcommand{\avg}[1]{\left\langle{#1}\right\rangle}

\begin{document}

\title{A Statistical Mechanics Model for the Emergence of Consensus}

\author{Giacomo Raffaelli and Matteo Marsili}

\affiliation{INFM-SISSA, via Beirut 2-4, Trieste I-34014, Italy and\\
Abdus Salam International Centre for Theoretical Physics\\
Strada Costiera 11, 34014 Trieste Italy}

\begin{abstract}
The statistical properties of pairwise majority voting over $S$
alternatives is analyzed in an infinite random population. We first
compute the probability that the majority is transitive (i.e. that if it prefers 
A to B to C, then it prefers A to C) and then study
the case of an interacting population. This is described by a
constrained multi-component random field Ising model whose ferromagnetic phase
describes the emergence of a strong transitive majority. We derive the
phase diagram, which is characterized by a tri-critical point and show
that, contrary to intuition, it may be more likely for an interacting
population to reach consensus on a number $S$ of alternatives when $S$
increases. This effect is due to the constraint imposed by transitivity on
voting behavior. Indeed if agents are allowed to express non transitive votes,
the agents' interaction may decrease considerably the probability 
of a transitive majority.
\end{abstract}

\maketitle

\section{Introduction}
Social choice and voting theory address the generic problem of how the individual 
preferences of $N$ agents over a number $S$ of alternatives can be 
aggregated into a social preference. This issue involves collective phenomena, such as
the emergence of a common opinion in a large population, which have attracted some
interest in statistical physics. For example, the voter \cite{Voter} 
and random field Ising models \cite{Galam} have been proposed to study how the 
vote's outcome between two alternatives is affected when voters influence each other.
In the case of two alternatives ($S=2$) the statistical mechanics of the majority vote
model has also been numerically studied on random graphs \cite{Moreira}.
In general, the framework for this kind of studies is
the statistical mechanics approach to socio-economic behavior 
\cite{Durlauf1,Schelling}, which stems from realizing that the 
emergence of ``macro-behavior"
can be the result of the interaction of many agents, each with their
own beliefs and expectations.

The majority rule can be naturally extended to $S>2$ alternatives by
considering the social preferences stemming from majority voting on
any pair of alternatives, i.e. pairwise majority rule (PMR). This
extension however is problematic, as observed back in 1785 by Marquis
de Condorcet \cite{Cond}. 
He observed that the PMR among three alternatives may
exhibit an irrational behavior, with the majority preferring
alternative A to B, B to C and C to A, even though each individual has
transitive preferences. These so-called Condorcet's cycles
may result in the impossibility to determine a
socially preferred alternative or a complete ranking of the
alternatives by pairwise majority voting (see also Ref. \cite{Meyer}
for a relation with statistical mechanics of dynamical systems). PMR
is not the only way to aggregate individual rankings into a social
preference \cite{Arrow,Dasgupta}. However the situation does not
improve much considering other rules. For example, the transitivity of
social preferences is recovered by resorting to voting rules
like Borda count, where each voter assigns a score to each alternative,
with high scores corresponding to preferred alternatives.
It turns out that these rules also violate
some other basic requirement. The basic desiderata of a social choice
rule are that it should be able to rank all alternatives for whatever
individual preferences ({\em unrestricted domain}), it should be {\em
transitive}, it should be {\em monotonous}, i.e. the social rank of an
alternative A cannot decrease when an individual promotes A to a
higher rank, and it should be {\em independent of irrelevant
alternatives}, i.e. the social preference between A and B cannot
depend on the preferences for other alternatives (independence of
irrelevant alternatives is important because it rules out the
possibility of manipulating the election's outcome by falsely
reporting individual preferences). For example, in plurality voting
each individual casts one vote for his top candidate and candidates
are ranked according to the number of votes they receive. This
satisfies all requirement but the last one, as vividly illustrated by
recent election outcomes \cite{Dasgupta}.

The discomfort of social scientists with the impossibility to find a reasonable voting 
rule has been formalized by Arrow's celebrated theorem \cite{Arrow}. 
This states that a social choice 
rule that satisfies all of the above requirements has to be {\em dictatorial},
that is there exists an agent -- the dictator -- such that the social preference between 
any two alternatives is the preference of that agent.

A way to circumvent the impasse of this result is to study the
properties of social choice rules on a restricted domain of possible
individual preferences. For example in politics, it may be reasonable
to rank all candidates from extreme left to extreme right. If the
preferences of each individual has a ``single peak'' when candidates
are ranked in this order (or any other order), then pairwise majority
is transitive. It has recently been shown that
pairwise majority turns out to be the rule which satisfies all
requirements in the largest domain \cite{Dasgupta}, thus suggesting
that pairwise majority is the best possible social choice rule.

In this paper we first try to quantify how good is majority rule by estimating the 
probability that pairwise majority yields a transitive preference relation in a typical 
case where individual preferences are drawn at random. 
This and closely related issues have 
been addressed by several authors \cite{Gilbaud,CW,Gehrlein}. 

Secondly, we study how the situation changes when agents influence each other.
In particular, as in the $S=2$ \cite{Voter,Galam}, we restrict to the relevant case where
the interaction arises from conformism~\cite{Schelling}. 
Basically conformism can stem from three different reasons \cite{Young}. 
It can be {\em pure} or {\em imitative}, because people simply want to be like others. It can be
due to the fact that in some cases conforming facilitates life ({\em
instrumental conformism}). Or it can be due to people deriving
information about the value of a choice from other people's behavior
({\em informational conformism}). In this light, our results may shed
light on a number of social phenomena, ranging from fashions or
fanaticism, where conformism may lead to the rise and spread of
broadly accepted systems of values, to the questions of how much
information should the agents share in order to achieve consensus on
$S$ items. At any rate, our discussion will focus on the
consequence of conformism on the collective behavior, 
without entering into details as to where this conformism stems
from. 

We show that the occurrence of a transitive social
choice on a number $S$ of alternatives for any choice of the
individual preferences, is related to the emergence of spontaneous
magnetization in a multi-component Ising model. We find a
phase diagram similar to that of the single component model
\cite{Aharony} with a ferromagnetic phase and a tricritical point
separating a line of second order phase transitions from a first order
one. The ferromagnetic state describes the convergence of a population
to a common and transitive preference ranking of alternatives, due to
social interaction.

Remarkably, we find that the ferromagnetic region expands as $S$
increases. Hence while without interaction the probability $P(S)$ of a
transitive majority vanishes rapidly as $S$ increases, if the
interaction strength is large enough, the probability of a transitive
majority increases with $S$ and it reaches one for $S$ large
enough. In other words, an interacting population may reach more
likely consensus when the complexity of the choice problem ($S$)
increases.

We finally contrast these findings with the case where agents need not
express a transitive vote (e.g. they may vote for A when pitted against
B, for B agains C and for C against A). This is useful because we find
that then the probability of finding a transitive majority is much
lower. In other words, individual coherence is crucial for conformism
to enforce a transitive social choice.

\section{Non-interacting population}

We shall first describe the behavior of a non-interacting population
and then move on to the interacting case. Let us consider a population of
$N$ individuals with preferences over a set of $S$ choices or
candidates. We shall mainly be interested in the limit $N\to\infty$ of
an infinite population. We limit attention to strict preferences,
i.e. we rule out the case where agents are indifferent between
items. Hence preference relations are equivalent to rankings of the
$S$ alternatives. It is convenient to represent rankings with
matrices $\hat\D_i$ for each agent $i=1,\ldots,N$, whose elements take
values $\D^{\a\b}_{i}=+1$ or $-1$ if $i$ prefers choice $\a$ to
$\b\neq \a$ or vice-versa, with $\a,\b=1,\ldots,S$. Notice that $\D^{\b\a}_{i}
=-\D^{\a\b}_{i}$. Let ${\cal R}$ be the set of matrices $\hat\D$ which
correspond to a transitive preference relation. Clearly the number of
such matrices equals the number $|{\cal R}|=S!$ of rankings of the $S$
alternatives. Hence not all the $2^{S(S-1)/2}$ possible asymmetric
matrices with binary elements $\D^{\a\b}_{i}=\pm 1$ correspond to
acceptable preference relations. For example, if $\D^{1,2} = \D^{2,3}
= \D^{3,1}$ then $\hat\D\not\in {\cal R}$. 
We use the term ranking to refer to matrices $\hat\Delta\in{\cal R}$
in order to avoid confusion later, when we will introduce preferences
over rankings, i.e. over elements of ${\cal R}$. We assume that each
agent $i$ is assigned a ranking $\hat\D_i$ drawn independently at
random from ${\cal R}$.

In order to compute the probability $P(S)$ that pairwise majority
yields a transitive preference relation, in the limit $N\to\infty$,
let us introduce the matrix
$\hat x =\frac{1}{\sqrt{N}}\sum_{i=1}^N \hat \D_i$.
The assumption on $\hat \D_i$ implies that the distribution of
$x^{\a\b}$ is Gaussian for $N\to\infty$ and it is hence completely
specified by the first two moments $\avg{x^{\a\b}}=0$ and
$\avg{x^{\a\b}x^{\c\d}}=\{\mat{G}^{-1}\}^{\a\b,\c\d}$ 
which is $0$ except for $\{\mat{G}^{-1}\}^{\a\b,\a\b}=1$,
$\{\mat{G}^{-1}\}^{\a\b,\a\d}=\{\mat{G}^{-1}\}^{\a\b,\c\b}=1/3$ and
$\{\mat{G}^{-1}\}^{\a\b,\c\a}=\{\mat{G}^{-1}\}^{\a\b,\b\d}=-1/3$,
where we have introduced the notation $\mat{M}$ for matrices with
elements $M^{ab,cd}$. The matrix $\mat{G}^{-1}$ can be inverted by a direct
computation, and we find that the matrix $\mat{G}$ has the same
structure of $\mat{G}^{-1}$ but with $\mat{G}^{ab,ab}=3\frac{S-1}{S+1}$,
$\mat{G}^{ab,ad}=-\frac{3}{S+1}=\mat{G}^{ab,cb}=-\mat{G}^{ab,bd}=
-\mat{G}^{ab,ca}$.

Let us first compute the probability $CW(S)$ that one of the alternatives,
is better than all the others. This means that there is a consensus
over the winner, while nothing is assumed for the relations
between the other choices. The preferred alternative is
known in social choice literature as the Condorcet winner, 
and much interest has been devoted to it, since
the presence of such a preferred alternative saves at least 
the possibility of electing a favorite choice.
$CW(S)$ is just the probability that $x^{1,\a}>0$ for all $\a>1$ multiplied by
$S$. In this way, we recover a known results \cite{CW}, which can be
conveniently casted in the form
\begin{equation}
CW(S)=S\sqrt{\frac{2}{\pi}}
\int_{-\infty}^{\infty} e^{-2 y^2+(S-1)\log [{\rm erfc}(y)/2]}.
\label{CW}
\end{equation}
Notice that $CW(S)$ is much larger than the na\"\i
ve guess $S/2^{S-1}$, derived assuming that $x^{\a\b}>0$
occurs with probability $1/2$ for all $\a\b$. Indeed asymptotic
expansion of Eq. (\ref{CW}) shows that 
\[
CW(S)\simeq 
\sqrt{\frac{\pi}{2}}\frac{\sqrt{\log
S}}{S}\left[1+O(1/\sqrt{\log S})\right]
\] 
decays extremely slowly for $S\gg 1$. 

The probability that the majority ranking is equal to the cardinal one
($1\succ 2\succ \ldots\succ S$) is given by the probability that
$x^{\a\b}>0$ for all $\a<\b$. This is only one of the $S!$ possible
orderings, then the probability of a transitive majority can be written as
\begin{equation}
P(S)=
S!\frac{[3/(2\pi)]^{\frac{S(S-1)}{4}}}{(S+1)^{\frac{S-1}{2}}}
\int_0^\infty \!d\hat x \exp 
\left[-\frac{1}{2}\hat x\cdot\mat{G}\cdot\hat x\right]
\label{PS}
\end{equation}
where $\int_0^\infty\!d\hat x\equiv\int_0^\infty \! dx_{1,2}\ldots
\int_0^\infty \! dx_{S-1,S}$ and we defined the product $\hat r\cdot\hat
q=\sum_{\a<\b} r^{\a\b}q^{\a\b}$ and its generalization to matrices
$\hat{r}\cdot\mat{M}\cdot\hat{q}=\sum_{a<b}\sum_{c<d}r^{ab}
M^{ab,cd}q^{cd}$. The normalization factor is computed from the
spectral analysis of $\mat{G}$ \cite{eigen}.

\begin{figure}[h]
  \centerline
      {
    \hbox
        {
          \psfig{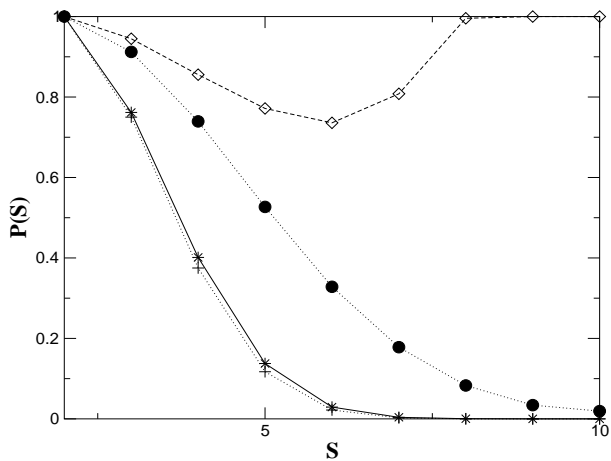}
        }
      }
      \caption{Probability $P(S)$ of a transitive majority
    ($\bullet$) compared to the na\"ive guess 
    $S!/2^{S(S-1)/2}$ ($+$).
    ($\diamond$) shows the case of an interacting
    population with $\beta=0.45$ and $\epsilon=0.8$,
    see Section \ref{iv}, $*$ show the same case 
    for the unconstrained case.
    }
      \label{figPS}
\end{figure}

We were not able to find a simpler form for this
probability. Fig. \ref{figPS} reports Montecarlo estimates of $P(S)$.
For $S=3$ we recover the result \cite{Gilbaud}
\begin{equation}
P(3)=CW(3)=\frac{3}{4}
+\frac{3}{2\pi}\sin^{-1}\frac{1}{3}\cong 0.91226\ldots
\end{equation}
 Again the na\"\i ve guess $P(S)\approx
S!/2^{S(S-1)/2}$ based on the fraction of acceptable rankings largely
underestimates this probability. This means that the collective
behavior of the majority hinges upon the (microscopic) transitivity of
individual rankings.

\vspace{0.7cm}

\section{Interacting voters}
\label{iv}
Let us now introduce interaction among voters. 
We assume that agents have an {\em a-priori} transitive preference
over the alternatives, specified by a ranking $\hat \Delta_i \in {\cal R}$.
We allow however agents to have a voting behavior which does
not necessarily reflect their {\em a-priori} ranking, that is,
we introduce a new matrix $\hat v_i$ such
that $v_i^{\a\b}=+1$ ($-1$) if agent $i$, in a context between $\a$
and $\b$, votes for $\a$ ($\b$). 
We will first study the case when $\hat v_i \in {\cal R}$, which corresponds
to agents having a rational voting behavior. This means
that even though an agent is influenced by others, she will maintain a
coherent choice behavior (transitivity). We will contrast this case
with that where the constraint on individual coherence $\hat v_i \in
{\cal R}$ is removed. 

To account for interaction, the matrix $\hat v_i$
depends not only on agents' preferences $\hat \Delta_i$,
but also on the interaction with other agents. Within
economic literature, this dependence is usually introduced
by means of an utility function $u_i$ which agents
tend to maximize.
Notice that this utility function represents
a preference over preferences (rankings).

Formally, this 
utility function depends both on an
idiosyncratic term $\hat\Delta_i \in {\cal R}$ describing the {\em a
priori} ranking, and on the behavior of other agents, $\hat
v_{-i}\equiv\{\hat v_j,~\forall j\neq i\}$, through the majority
matrix
\begin{equation}
\hat m=\frac{1}{N}\sum_{i=1}^N \hat v_i.
\label{m}
\end{equation}
More precisely, we define an utility function
\begin{equation}
u_i(\hat v_i,\hat v_{-i})=
(1-\e)\hat\D_i \cdot\hat v_i+\e \hat m\cdot \hat v_i.
\label{ui}
\end{equation}
where the last term captures conformism as a diffuse preference for
aligning to the majority \cite{Schelling,Young}. 
For $\e=0$ maximal
utility in Eq. (\ref{ui}) is attained when agents vote as prescribed
by their {\em a priori} rankings, i.e. $\hat v_i=\hat \D_i$ $\forall
i$.
On the contrary, for $\e=1$ agents totally disregard their
rankings and align on the same ranking $\hat v_i=\hat m$ $\forall i$,
which can be any of the $S!$ possible ones.

Let us characterize the possible stable states, i.e. the Nash
equilibria of the game defined by the payoffs of Eq. (\ref{ui}). These
are states $\hat v^*_i$ such that each agent has no incentives to
change his behavior, if others stick to theirs,
i.e. $u_i(v_i,v_{-i}^*)\le u_i(v_i^*,v_{-i}^*)$ for all $i$.
The {\em random} state $\hat v_i^*=\hat \D_i$ is (almost surely) a
Nash equilibrium $\forall \e<1$, because the payoff of aligning to the
majority $\hat m=\hat x/\sqrt{N}$ is negligible with respect to that
of voting according to own ranking $\hat \Delta_i$. Then we have $u_i(\hat
\D_i,\hat \D_{-i})=\frac{S(S-1)}{2}[1-\e+\e O(1/\sqrt{N})]$.  This
Nash equilibrium is characterized by a majority which is not
necessarily transitive, i.e. which is transitive with probability
$P(S)<1$ for $N\gg 1$.

Also {\em polarized} states with $\hat v_i=\hat m$ for all $i$ are
Nash equilibria for $\e> 1/2$. Indeed, with some abuse of notation,
when all agents take $\hat v_j=\hat m$ for some $\hat m$, 
agent $i$ receives an utility
$u_i(\hat m,\hat m)=(1-\e)\hat\D_i\cdot\hat m+
\frac{S(S-1)}{2}\e$. The agents who are worse off are those with
$\hat \D_i=-\hat m$ for whom $u_i(\hat m,\hat
m)=\frac{S(S-1)}{2}[2\e-1]
\cong -u_i(\hat \D_i,\hat m)+O(1/N)$. Then as long as $\e> 1/2$, even
agents with 
$\hat\D_i=-\hat m$ will not profit from abandoning the
majority. Therefore $\hat v_i=\hat m$ for all $i$ is a Nash
equilibrium. Notice that whether the majority is transitive ($\hat m\in {\cal R}$) 
or not depends on whether agents express transitive preferences ($\hat v_i\in 
{\cal R}$) or not. In the former case the majority will be transitive whereas
if non transitive voting is allowed there is no need to have $\hat m\in {\cal
  R}$ and there are $2^{S(S-1)/2}$ possible polarized Nash equilibria. Only
in $S!$ of them the majority is transitive (i.e. when $\hat m\in {\cal
  R}$).

It is easy to check that there are no other Nash equilibria. 
Summarizing, for $\e>1/2$ there are many Nash equilibria. Depending
on the dynamics by which agents adjust their voting behavior one or
the other of these states will be selected. 

\section{Statistical Mechanics of interacting voters}

Strict utility maximization leads to the presence of multiple equilibria, 
leaving open the issue of which equilibrium will the population select. 
It is useful to generalize the strict utility
maximization into a stochastic choice behavior which allows for
mistakes (or experimentation) with a certain probability
\cite{KMR}. This on one side may be realistic in modelling many 
socio-economic phenomena \cite{Durlauf1,Durlauf,Young}. On the other 
this rescues the uniqueness of the solution, in terms of
the probability of occurrence of a given state $\{\hat v_i\}$, under
some ergodicity hypothesis. 
Here, as in \cite{Durlauf}, we assume that agents have the
following probabilistic choice behavior: agents are asynchronously
given the possibility to revise their voting behavior. When agent $i$
has a revision opportunity, he picks a voting profile $\hat w$ ($\in {\cal
R}$ when voters are rational) with probability
\begin{equation}
P\{\hat v_i=\hat w\}=Z_i^{-1} e^{\beta u_i(\hat w,\hat v_{-i})}
\label{Pv}
\end{equation}
where  $Z_i$ is a normalization constant. Without entering into
details, for which we refer to Ref. \cite{Durlauf}, let us
mention that Eq. (\ref{Pv}) does not necessarily assume that agents
randomize their behavior on purpose. It models also cases where agents
maximize a random utility with a deterministic term $u_i$ and a random
component. Then the parameter $\beta$ is related to the
degree of uncertainty (of the modeler) on the utility
function \cite{other_dyn}. 

When agent $i$ revises his choice 
the utility difference $\delta u_i=u_i(\hat v_i,\hat v_{-i})-u_i(\hat
v_i',\hat v_{-i})$ for a change $\hat v_i\to \hat v_i'$ is equal to
the corresponding difference in $-H$, where
\begin{equation}
H\{\hat v_i\}=-(1-\e)\sum_{i=1}^N\hat\D_i \cdot\hat v_i-
\frac{\e}{2N}\sum_{i, j=1}^N  \hat v_j\cdot \hat v_i.
\label{H}
\end{equation}
hence in the long run, the state of the population
will be described by the Gibbs measure $e^{-\beta H}$
because the dynamics based on
Eq. (\ref{Pv}) satisfies detailed balance with the Gibbs measure. 

$H$ in Eq. (\ref{H}) is the Hamiltonian of a multi-component random field
Ising model (RFIM) where each component $v_i^{ab}$ with $a<b$ is
a component of the spin, $\hat\Delta_i$ represents the random field
and the term 
$\frac{\e}{2N}\sum_{i, j=1}^N  \hat v_j\cdot \hat v_i$ 
is a mean field interaction. 
Indeed $\hat v_i$ has $S(S-1)/2$ components which take
values $v_i^{\a\b}=\pm 1$. The peculiarity of this model is that
the components of the fields $\hat \Delta$ are not independent. 
Indeed not all the $2^{S(S-1)/2}$
values of $\hat \D_i$ are possible but only those $\hat \D_i\in {\cal
R}$, which are $S!$. The same applies to the spin components $\hat v_i$
when rational voting behavior is imposed. Were it not for this
constraint, the model would just correspond to a collection of
$S(S-1)/2$ uncoupled RFIM. 

The statistical
mechanics approach of the RFIM
\cite{Schneider-Pytte,Aharony} can easily be generalized to the
present case. The partition function can be written as
\begin{equation}\label{Z}
Z(\beta)={\rm Tr}_{\{\hat v_i\}}e^{-\beta H}=\int d\hat m e^{-N\beta f(\hat m)}
\end{equation}
where the trace ${\rm Tr}_{\{\hat v_i\}}$ over spins runs on all $\hat
v_i\in {\cal R}$ when voting behavior is
rational, or over all $\hat v_i$ otherwise.
The free energy $f(\hat m)$ is given by
\begin{equation}
f(\hat m)=\frac{\e}{2}\hat m^2-\frac{1}{N\beta}\sum_{i=1}^N
\log\left[\sum_{\hat v} e^{\beta[(1-\e)\hat\D_i+\e \hat
      m]\cdot\hat v}\right]
\end{equation}
where once again the sum over the $\hat v_i$ runs inside
$\cal R$ if agents are rational, or is not limited
otherwise.
It is evident that $f$ is self averaging. Hence in the limit
$N\to\infty$ we can replace $\frac{1}{N}\sum_i\ldots$ with the
expected value $\frac{1}{S!}\sum_{\hat\D\in{\cal R}}\ldots\equiv 
\avg{\ldots}_{\D}$ on $\hat\D_i$. It is also clear that the integral
over $\hat m$ of Eq. (\ref{Z}) in this
limit is dominated by the saddle point value
\begin{equation}
\hat m=\left\langle
\frac{\sum_{\hat v} \hat v e^{\beta[(1-\e)\hat\D+\e \hat
      m]\cdot\hat v}}
{\sum_{\hat v} e^{\beta[(1-\e)\hat\D+\e \hat
      m]\cdot\hat v}}
\right\rangle_{\D}
\label{magnetization}
\end{equation}

\begin{figure}[h]
\centerline{
\hbox{
\psfig{figure=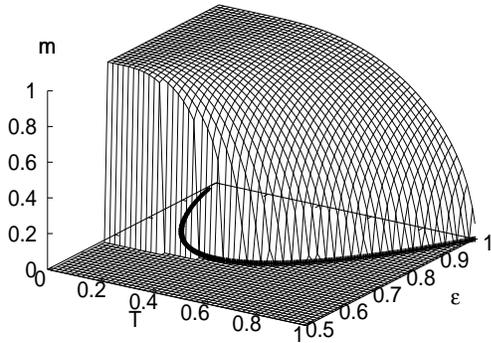,height=6cm,width=8cm}
}
}
  \caption{Phase diagram for $S=2$. The  plot shows the magnetization,
    while on the bottom we have drawn the line that marks the instability
    of the paramagnetic solution.}
  \label{fig0}
\end{figure}

This equation can be solved from direct iteration
and shows that for large enough values of
$\beta>\beta_c$ there is a transition, as $\e$ increases, from a 
paramagnetic state
with $\hat m=0$ to a polarized (ferromagnetic) state where $\hat m\neq 0$. 
In Fig. (\ref{fig0}) we plot the result of such iterative solution for 
the RFIM case, i.e. when $S=2$. Since for some values $T, \epsilon$ 
both the ferromagnetic and the paramagnetic state can be stable, 
we have solved for the magnetization $\hat m$ starting both
from a $\hat m = 0$ and from $\hat m = \hat 1$ states. 
Then we selected the correct equilibrium state by comparing
the free energy of the different solutions.  
The stability of the paramagnetic solution $\hat m=0$ can be inferred
from the expansion of Eq. (\ref{magnetization}) around $\hat m=0$,
which reads 
\begin{equation}
\hat m=\beta\e\mat{J}\cdot\hat m+O(\hat m^3)
\end{equation}
where
\begin{equation}
J^{\a\b,\c\d}=\left\langle\avg{v^{ab}v^{c,d}|\hat\D}_v-
\avg{v^{ab}|\hat\D}_v\avg{v^{c,d}|\hat\D}_v\right\rangle_\Delta.
\label{J}
\end{equation}
Here averages $\avg{\ldots|\hat\D}_v$ over $\hat v$ are taken with the
distribution 
\begin{equation}
\label{vdistr}
P(\hat v|\hat \D)=\frac{e^{\beta(1-\e)\hat\D\cdot\hat v}}
{\sum_{\hat u} e^{\beta(1-\e)\hat\D\cdot\hat u}}.
\end{equation}
When the largest eigenvalue $\Lambda$ of $\beta\e\mat{J}$ is larger
than one, the paramagnetic solution $\hat m=0$ is unstable and only
the polarized solution $\hat m\neq 0$ is possible.
In Fig.\ref{fig0} the line that marks the region of 
instability of the paramagnetic solution 
is plotted at the bottom.

\subsection{Constrained case, $\hat v_i \in{\cal R}$}

\begin{figure}[h]
\centerline{
\hbox{
\psfig{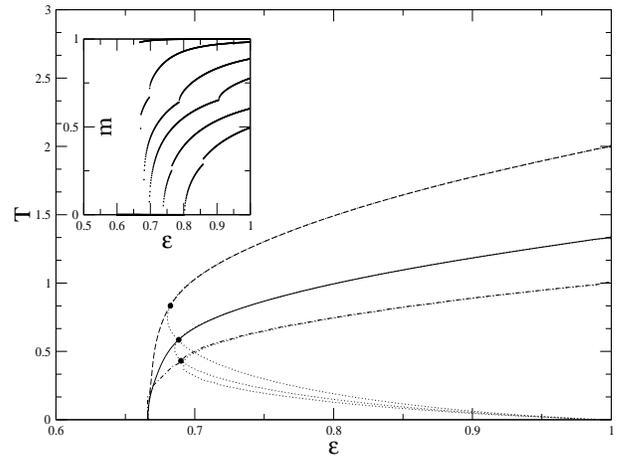}
}
}
  \caption{Phase diagram for $S=2,~3$ and $5$ (dot-dashed, straight
    and dashed lines). Dotted lines mark the region where the $\hat m=0$ phase
  is unstable. These meet the lines across which the
  transition takes place, at the tricritical point ($\bullet$). Inset:
  magnetization for $1/\beta=0.25,\,0.5,\ldots,1.75$ and $S=5$ as a
  function of $\e$.}
  \label{phase}
\end{figure}

Here both the individual {\em a-priori}
rankings $\hat \Delta_i$ and the voting behavior
$\hat v_i$ of each agent are transitive. 
Results for the numerical iteration of Eq. (\ref{magnetization})
are shown in the inset of Fig.\ref{phase} for
different values of $\beta$ and for $S=5$. 
Fig. \ref{phase} shows the phase diagram for $S=2,~3$ and for 
$S=5$. 
The transition from the paramagnetic
phase to the ferromagnetic one is continuous for intermediate values of $\beta$
($\beta_t<\beta<\beta_c$) but becomes discontinuous when
$\beta>\beta_t$. The transition point $\beta_t$ ($\bullet$) generalizes
the tricritical point of the RFIM \cite{Aharony}
($S=2$).

\vspace{1.0cm}

 The condition
$\Lambda=1$ on the largest eigenvalue $\Lambda$ of $\beta\epsilon{\cal
   J}$ reproduces the second order transition line. The line
$\Lambda=1$ continues beyond the tricritical point and it marks the border of
the region where the paramagnetic solution $\hat m=0$ is unstable
(dotted line in Fig. \ref{phase}). Below the lower branch of
the $\Lambda=1$ line the paramagnetic solution is locally stable
but it is not the most probable. Indeed the polarized state $\hat m^*$
which is the non-trivial solution of Eq. (\ref{magnetization}) has a
lower free energy $f(\hat m^*)< f(0)$. Still in numerical simulation
the state $\hat m=0$ can persist for a very long time in this
region. The polarized solution $\hat m^*$ becomes metastable and then
disappears to the left of the transition line in Fig. \ref{phase}.

With respect to the dependence on $S$ of the phase diagram, we observe
that at $\beta\to\infty$ the phase transition takes place at $\e=2/3$
independent of $S$. At the other extreme, for $\e=1$ we find that
\cite{proof} $\mat{J}=\mat{G}^{-1}$. The
largest eigenvalue of $\beta\mat{J}$ is thus \cite{eigen}
$\Lambda=\beta\frac{S+1}{3}$ and the condition $\Lambda=1$ implies
that
\begin{equation}
\beta_c(\e=1)=\frac{3}{S+1}
\label{betac}
\end{equation}
Hence as $S$ increases the region where the polarized phase is stable
becomes larger and larger. In other words it
becomes more and more easy for a population of agents who influence
each other to become polarized on the same opinion.  This is somewhat
at odd with na\"\i ve expectation, because as $S$ increases the
complexity of the choice problem also increases and reaching consensus
becomes more difficult. Indeed the probability $P(S)$ to find
consensus on $S$ choices in a random population drops very rapidly to
zero as $S$ increases. Nevertheless, the effects of interaction toward
conformism becomes stronger. We attribute this to the fact that for
large $S$ the
fraction of allowed spin configurations $\hat v\in {\cal R}$ is
greatly reduced, thus inducing a strong interaction among
the different spin components. This results in the fact that ordering
becomes easier and easier when $S$ increases.

\subsection{Unconstrained case}
Here the constraint $\hat v_i \in{\cal R}$ is not imposed, 
while we keep $\hat \D \in \cal R$. This means
that an agent can be influenced by other agents'
preferences to the point of picking an intransitive
preference. In this case all the traces over the $\hat v_i$ in the above equations 
can be computed component-wise, independently, as in
a multi-component random field Ising model.
A direct computation of the matrix 
$J^{\a\b,\c\d}$ is possible, and yields 
\begin{equation}
J^{\a\b,\c\d} = \delta_{\a\c}\delta_{\b\d}\left[
1-\tanh^2(\beta(1-\epsilon)\right].
\end{equation}
Notice that, for any $\beta$ and $\epsilon$,
the maximum eigenvalue $\Lambda=\beta\epsilon [
1-\tanh^2(\beta(1-\epsilon)]$ of the matrix $\beta\epsilon\cal J$
is independent of $S$ and it coincides with that of the RFIM ($S=2$). 
Hence the phase diagram is that of the RFIM for all $S\ge 2$. 
The different spin components behave independently. The correlation 
induced by the constraints on the {\em a-priori} preferences 
$\D-i\in{\cal R}$ does not influence the thermodynamics properties.
Note that for $\epsilon \to 1$ the condition $\Lambda = 1$ 
implies $\beta = 1$ and for $\beta\to\infty$ the phase transition takes 
place at $\e=2/3$, independent of $S$. 

\section{$P(S)$ with interacting voters}

The main result of the previous section, that is, 
the fact that ordering becomes easier as $S$ increases
when rational voting behavior is assumed for each agent,
has interesting effects on the probability of finding a 
transitive majority. To investigate this, we analyze
the probability
$P_{\beta,\e}(S)$ of a transitive majority in an interacting
population. The calculation is a generalization
of the one presented for the non-interacting population.
Let 
\[
\z=\frac{1}{\sqrt{N}}\sumi \vi.
\]
We want to compute, at a fixed $\epsilon$ and $\beta$, the probability distribution of $\z$.
Keeping fixed the realization of the disorder $\Di$, this is given by

\begin{eqnarray*}
&P \left( \hat z|\{\hat \D_i\}\right) = 
{\cal N}
\Tr_{\hat v_i}
e^{-\beta {\cal H}\vi}
\delta\left(\z-\frac{1}{\sqrt{N}}\sumi \vi\right)\\
&={\cal N} e^{\frac{\beta\epsilon}{2}\z\cdot\z}\int d\lam e^{i\lam\cdot\z}
\prod_{i=1}^N \Tr_{\vi} e^{[\beta(1-\epsilon)\Di-i\lam/\sqrt{N}]\cdot\vi}
\end{eqnarray*}
now the term $\lam/\sqrt{N}$ is small compared to the other one and we can expand it

\begin{eqnarray*}\label{tri}
&&  \Tr_{\vi} e^{[\beta(1-\epsilon)\Di-i\lam/\sqrt{N}]\cdot\vi}=\\
&=&\Tr_{\vi} e^{\beta(1-\epsilon)\Di\cdot\vi}\left[1
  -\frac{i}{\sqrt{N}}\lam\cdot\vi-\frac{1}{2N}(\lam\cdot\vi)^2+\ldots\right]\\
&=&  \Tr_{\vi} e^{\beta(1-\epsilon)\Di\cdot\vi}\left[1-\frac{i}{\sqrt{N}}\lam\cdot\avg{\vv|\Di}-\right.\\
&-&\left.\frac{1}{2N}\sum_{ab,cd}\lambda^{ab}\lambda^{cd}\avg{\vv^{ab}\vv^{cd}|\Di}+\ldots\right]
\end{eqnarray*}
where, again, averages over the $\hat v$ are taken with the distribution
(\ref{vdistr}).
 The factor $Z_i=\Tr_{\vi} e^{\beta(1-\epsilon)\Di\cdot\vi}$ can 
be absorbed in the normalization constant, so that if we re-exponentiate the terms, we find
\begin{eqnarray*}
&&\Tr_{\vi} e^{[\beta(1-\epsilon)\Di-i\lam/\sqrt{N}]\cdot\vi} \cong  \\
&&\cong Z_i e^{-\frac{i}{\sqrt{N}}\lam\cdot\avg{\vv|\Di} -\frac{1}{2N}
\sum_{ab,cd}\lambda^{ab}\lambda^{cd}{\cal J}^{ab,cd}}
\end{eqnarray*}
This gives
\begin{eqnarray*}
P\left(\z|\{\Di\}\right)&=&{\cal N}' e^{\frac{\beta\epsilon}{2}\z\cdot\z}
\int d\lam e^{i\lam\cdot(\z-\y)-\frac{1}{2}\lam\cdot{\cal J}\cdot\lam}\\
&=&{\cal N}" e^{\frac{\beta\epsilon}{2}\z\cdot\z-\frac{1}{2}(\z-\y)
\cdot{\cal J}^{-1}\cdot(\z-\y)}\\
&=&{\cal N}" e^{-\frac{1}{2}\z\cdot[{\cal J}^{-1}-\beta\epsilon{\cal I}]
\cdot\z+\z\cdot{\cal J}^{-1}\cdot\y-
\frac{1}{2}\y\cdot{\cal J}^{-1}\cdot\y}\\
\hbox{where}~&~&~~\y=\frac{1}{\sqrt{N}}\sumi \avg{\vv|\Di}
\end{eqnarray*}
and ${\cal J}$ given by Eq. (\ref{J}).
Now one needs to take the average over $P(\y)$. In general this is a 
Gaussian distribution

\begin{equation}\label{Py}
  P(\y)\propto e^{-\frac{1}{2}\y\cdot {\cal A}\cdot \y}
\end{equation}
and, considering the $\y$ dependence of the normalization ${\cal N}"$
\[
{\cal N}"\propto e^{\frac{1}{2}\y\cdot{\cal J}^{-1}\cdot\y-\frac{1}{2}
\y\cdot\frac{1}{{\cal J}-\beta\epsilon {\cal J}^2} \cdot\y}
\]
we get
\begin{equation}\label{Px}
  P(\z)\propto e^{-\frac{1}{2}\z\cdot{\cal K}\cdot \z}
\end{equation}
where

\begin{equation}\label{kk}
  {\cal K}={\cal J}^{-1}-\beta\epsilon {\cal I}-
  \frac{1}{{\cal J}{\cal A}{\cal J}+\frac{1}{{\cal J}^{-1}-
      \beta\epsilon{\cal I}}}
 \end{equation}

As before, this probability can be computed to the desired level of
accuracy with the Montecarlo method. 

\subsection{Constrained case}

When $\hat v_i\in {\cal R}$ we have 

\begin{equation}\label{Acon}
  \left\{{\cal A}^{-1}\right\}^{ab,cd}=\left\langle \langle v^{ab}|\D\rangle
  \langle v^{cd}|\D\rangle\right\rangle_{\Delta}.
\end{equation}

Fig. \ref{figPS} ($\diamond$) shows that the resulting
$P_{\beta,\e}(S)$ may exhibit a non-monotonic behavior with $S$: first
it decreases as $P(S)$ and then, as the point $(\beta,\e)$ approaches
the phase transition line it starts increasing. If $\e>2/3$, there
is a value $S^*$ beyond which the system enters in the polarized phase
and $P_{\beta,\e}(S)=1$ $\forall S\ge S^*$. 

\subsection{Unconstrained case}

In this case $\avg{\vv|\Di}=t \Di$ where we introduce the 
shorthand $t=\tanh [\beta(1-\epsilon)]$. Then
${\cal A}={\cal G}/t^2$ or

\begin{equation}\label{Py}
  P(\y)\propto e^{-\frac{1}{2t^2}\y\cdot {\cal G}\cdot \y}
\end{equation}
in addition

\begin{equation}\label{Junc}
 {\cal J}=(1-t^2){\cal I}
\end{equation}
hence setting $f = 1-\beta\epsilon(1-t^2)$
\begin{equation}\label{kk}
  {\cal K}=\left[\frac{1}{1-t^2}-\beta\epsilon\right]{\cal I}-
  \frac{t^2}{1-t^2}\frac{f}{t^2{\cal I}+f(1-t^2){\cal G}}.
\end{equation}

The behavior of the probability can be understood in some interesting 
limits. For $\beta\to\infty$ we get
\[
{\cal K}\simeq {\cal G}+O(1-t^2)
\]
which simply states that as the temperature goes to zero
the probability reduces to that of the constrained case, as it should. 
Note that ${\cal K}\to {\cal G}$ also as we approach the critical line 
where $1-\beta\epsilon(1-t^2)\to 0$. 

Instead for $\epsilon\to 0$ we have
\[
{\cal K}\to \frac{1}{1-t^2}\left[{\cal I}-\frac{1}{{\cal I}+(t^{-2}-1){\cal G}}\right]
\]
The high $T$ limit $\beta\to 0$ reads
\[
{\cal K}\simeq {\cal I}-\beta^2{\cal G}^{-1}+\ldots
\]
that is, since the matrix ${\cal K}$ is diagonal 
the probability of finding a transitive majority drops to the 
trivial one, namely $S! 2^{-S(S-1)/2}$. So without the constraint
of rational voting the probability of a transitive outcome 
can be greatly reduced.
Again, Monte Carlo simulations are shown in Fig.\ref{PS} ($*$). Note the
marked decrease of the probability of finding a transitive majority
with respect to the constrained and to the non-interacting case.

\section{Conclusions}

In conclusion we have studied the properties of pairwise majority
voting in random populations. We have computed the probability that
pairwise majority is transitive when there is no interaction and found
that it decreases rapidly with $S$, even though less
rapidly than one would naively guess. Then we have shown that the
properties of pairwise majority in a random interacting population are
related to the properties of a multi-component RFIM, whith 
a constraint on the components which reflects the transitivity
of individual preferences. This model can be
solved exactly and features a ferromagnetic phase where the
population reaches a consensus (i.e. a transitive majority) with
probability one. As to the dependance on the number of voters,
we find that the ferromagnetic phase gets larger and larger as $S$
increases, meaning that consensus is reached more easily
when the complexity of the problem (i. e. the number of alternatives)
is large enough. 

With respect to the case when rational voting behavior
is not imposed, we note the strikingly different
effect that intereaction can have, dependant on
how this interaction is introduced.
In fact, if we impose a transitive voting behavior,
the probability to find a transitive
majority is increased, while relaxing this constraint
can result in a decrease of this probability.


\begin{thebibliography}{200}
\bibitem{other_dyn} A similar picture was found with an adaptive
  dynamics where agent learn about their best choice over time.
\bibitem{proof} With $\e=1$, $J^{\a\b,\c\d}=\frac{1}{S!}\sum_{\hat
v\in {\cal R}}v^{ab}v^{c,d}$. So $J^{ab,ab}=1$ and $J^{ab,cd}=0$ by
symmetry if $a\neq c,d$ and $b\neq c,d$. Furthermore $J^{ab,ad}$
depends only on the relative ordering between $a,b$ and $d$ in the
permutation $\hat{v}$. The permutations where $a$ is
between $b$ and $d$, which are $1/3$, give $v^{ab}v^{ad}=-1$, whereas
the remaining $\hat v$ give $v^{ab}v^{ad}=1$. Hence 
$J^{ab,ad}=\frac{\beta}{3}$. Likewise we find
$J^{ab,cb}=-J^{ab,ca}=-J^{ab,bd}=\frac{\beta}{3}$.
\bibitem{eigen} The matrix $\mat{G}$ has $S-1$ eigenvectors of the
form $z^{ab}_{1,k}=\delta_{a,k-1}{\rm sign}(b-k+1)$ with $k=2,\ldots,
S$, and eigenvalue $\lambda=3/(S+1)$. This can be verified by direct
calculation. Note that the vectors $z^{ab}_k$ are not orthogonal, but
are linearly independent. Direct substitution shows that all vectors
of the form $z^{ab}_{j,k}=\delta_{a,1}(\delta_{b,j}-\delta_{b,k})+
\delta_{a,j}\delta_{b,k}$ are also eigenvectors of $\mat{G}$ for
$1<j<k$, with eigenvalue $\lambda=3$.  Then $\lambda=3$ has degeneracy
$(S-1)(S-2)/2$. The set of $S(S-1)/2$ linearly independent vectors
$z^{ab}_{j,k}$ with $1\le j<k\le S$ allows us to build a complete
orthonormal basis of eigenvectors and to compute $\det\mat{G}$.
\end{thebibliography}
\end{document}